\documentclass[aps,prb,reprint,amsfont,amsmath,amssymb,floatfix]{revtex4-2}
\usepackage{graphicx,bm,xcolor}
\usepackage[normalem]{ulem}
\usepackage{hyperref}
\hypersetup{colorlinks=true,allcolors=blue}

\begin{document}

\title{Shear Bernstein modes in a two-dimensional electron liquid}
\author{A.~N.~Afanasiev}
\email{afanasiev.an@mail.ru}
\author{P.~S.~Alekseev, A.~A.~Greshnov, M.~A.~Semina}
\affiliation{Ioffe Institute, St.~Petersburg 194021, Russia}

\begin{abstract}

Bernstein modes are formed as a result of non-local coupling of collective excitations and cyclotron harmonics in magnetized plasma. In degenerate solid state plasma they are typically associated with magnetoplasmons. A different type of Bernstein modes arises in two-dimensional electron liquid at sufficiently strong quasiparticle interaction. We consider Bernstein modes originating from coupling between quasiparticle cyclotron harmonics and shear magnetosound waves. The latter may be responsible for the giant peak in radio-frequency photoresistance observed in high-quality GaAs quantum wells. Using Landau-Silin kinetic equation with an arbitrary strength of the interparticle Landau interaction, we trace the reconstruction of Bernstein mode spectrum in high-quality 2D electron systems across the crossover between weakly interacting degenerate electron gas and the correlated electron liquid. Sensitivity of Bernstein modes to the strength of quasiparticle interaction allows one to use them for spectroscopy of Landau interaction function in the electron Fermi liquids.

\end{abstract}

\maketitle

\section{Introduction}
Recent demonstration of hydrodynamic electron transport~\cite{Polini2020,Narozhny2022} in ultrapure samples of graphene~\cite{Bandurin2016,Crossno2016,Sulpizio2019,Ku2019,Kumar2022}, 2D metals~\cite{Moll2016,Steinberg2022} and semiconductor heterostructures~\cite{Alekseev2016,Gusev2018,Gusev2020,Wang2023} has marked the practical realization of pristine two-dimensional electron liquid (2DEL), which dynamics is governed solely by the electron-electron interaction while the effects of disorder and electron-phonon interaction are very minor. Experimental studies of high-frequency response of hydrodynamic samples, namely their photoresistance in classically strong magnetic fields, revealed non-sinusoidal shape of microwave-induced resistance oscillations (MIRO)~\cite{Smet2005,Dai2010,Hatke2011,Herrmann2016} and abnormally strong absorption at cyclotron resonance overtones~\cite{Dai2010,Hatke2011,Herrmann2016,Bandurin2022,Du2022}. The latter effect is especially pronounced in GaAs quantum wells at the second cyclotron harmonic~\cite{Dai2010,Hatke2011,Herrmann2016,Du2022}.

High-frequency properties of electron liquid are typically associated with the collisionless collective modes~\cite{Pines_Nozieres_2018,Platzman1973}, and the effect of giant photoresponse was attributed to excitation of the two different types of waves. First of them~\cite{Bandurin2022,Volkov2014,Kapralov2022} are Bernstein modes (BMs)~\cite{Sitenko1956,Bernstein1958} in 2D degenerate electron gas (2DEG)~\cite{Chiu1974,Kukushkin2016}. At the wave vectors comparable to the inverse cyclotron radius, magnetoplasmon and single particle excitations are coupled due to non-uniformity of the self-consistent electrostatic field of the wave at the cyclotron orbit. As a result, long wavelength magnetoplasmon dispersion splits into series of BM branches between the sequential cyclotron harmonics.

The second explanation~\cite{Alekseev2019_PRL,Alekseev2019_SC} of the giant photoresponse relies on viscoelasticity of 2DEL~\cite{Conti1999}. At sufficiently high frequencies, $\omega \gg \nu_2$ (here $\nu_2$ is the inverse shear stress relaxation time determined by the electron-electron collisions), the viscosity coefficient becomes purely imaginary and proportional to the elastic shear modulus. Behavior of the electron liquid is similar to amorphous solid in this frequency domain: shear stress density $\hat{\Pi}_{\rm sh}({\bf r},t)$ oscillates in phase with particle current density ${\bf j}({\bf r},t)$ and thus can propagate in the form of shear sound, in contrast to attenuation in hydrodynamic regime $\omega\ll \nu_2$~\cite{Landau_V7}. From the microscopic point of view, shear wave in electron liquid is represented by transverse zero sound~\cite{Landau_FL:ZS}. It emerges in 2DEL with relatively strong dipole part of the short-range Landau interaction, when quasiparticle mass $m=(1+F_1)m^{(0)}$~\cite{Renorm_Note} exceeds the threshold value of $2m^{(0)}$~\cite{Khoo2019,Khoo2020,Valentinis2021} (here $F_1$ is the spin-symmetric dipole Landau parameter and $m^{(0)}$ is the effective electron mass before the interaction is turned on). The required degree of mass renormalization has been observed in high purity semiconductor heterostructures with low carrier concentration~\cite{Shashkin2002,Pudalov2002,Anissimova2006,Gokmen2010,Kuntsevich2015,Falson2015,Melnikov2023}.

In external magnetic field, shear stress density is determined by the two viscosity coefficients: ordinary (even) $\eta_s(\omega)$ and Hall (odd) $\eta_{H}(\omega)$~\cite{Avron1995,Avron1998}, both having resonance at the renormalized second cyclotron harmonic~\cite{Alekseev2018}. This viscoelastic resonance manifests itself via reconstruction of the collective excitation spectrum at long wavelengths. Besides the conventional magnetoplasmon, spectrum consists of the shear magnetosound mode~\cite{Alekseev2019_PRL,Alekseev2019_SC}, which leads to the giant peak in photoresistance at the second cyclotron harmonic~\cite{Dai2010,Hatke2011,Herrmann2016,Du2022,Alekseev2019_PRL,Alekseev_Alekseeva_2022}. However, the viscoelastic model is valid~\cite{Alekseev2019_SC} only in the long wavelength domain for strongly-interacting 2DEL, while the quasiparticle interaction strength is arbitrary and dependent on $r_s$ in realistic GaAs structures.

In this Letter we consider the evolution of collective mode spectrum of magnetized 2DEL with the increase of quasiparticle interaction strength. Using Landau-Silin kinetic equation~\cite{Silin1958,Platzman1973,Pines_Nozieres_2018} we show that the conventional BM pattern of 2DEG is reconstructed at strong quasiparticle mass renormalization (high $F_1$). The emergent shear magnetosound mode couples with the quasiparticle excitations and forms another type of BMs. Renormalized second cyclotron resonance harmonic becomes the main one and the resulting collective excitation spectrum consists of the two independent series of BMs. The {\it conventional BMs} are related to magnetoplasmon and characterized by the longitudinal current polarization, while the coexisting {\it shear BMs} related to shear magnetosound carry the transverse current density.


\section{Long wavelength excitations in magnetized viscoelastic 2DEL}
\begin{figure}[t!]
    \includegraphics[width=0.49\textwidth]{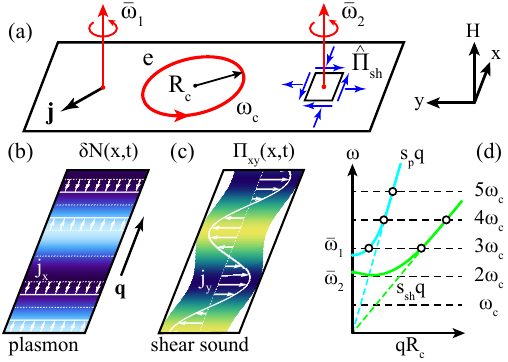}
    \caption{Formation of Bernstein modes in 2DEL with strong dipole quasiparticle interaction; a) cyclotron rotation of current density, shear stress density and individual quasiparticles (marked by red circles) in magnetic field; b) and c) propagation of plasmon and shear sound in 2DEL: quasiparticle and shear stress densities in the waves are marked by color, white arrows denote current density fields, shape of the colored region correspond to the typical displacement of the finite liquid element under wave propagation; d) long wavelength dispersions of magnetoplasmon (cyan) and shear magnetosound (light green) in gated 2DEL and their counterparts at zero magnetic field (dashed lines), white circles mark the positions of anti-crossings with single particle excitations at $n\omega_c$ leading to Bernstein modes.}
    \label{Fig:Scheme}
\end{figure}

Single-particle excitations of magnetized 2DEL are represented by quasiparticle transitions to higher Landau levels above the topmost occupied one in the ground state. At large filling factors, they are characterized by harmonics of the renormalized cyclotron frequency $\omega_c=eH/mc$~\cite{Renorm_Note} and the wave packets describing the excited quasiparticles are localized at circular orbits with cyclotron radius $R_c=v_F/\omega_c$ up to the Fermi wavelength (here $v_F$ is the Fermi velocity). In contrast to the weakly-interacting 2DEG case, long wavelength cyclotron resonance frequencies $\bar{\omega}_n$ in magnetized 2DEL differ from the quasiparticle cyclotron harmonics $n\omega_c$. In 2DEL, local cyclotron rotation of macroscopic quantities proportional to the angular harmonics of non-equilibrium distribution function (e.g. current and shear  stress densities) is affected by quasiparticle backflow~\cite{Pines_Nozieres_2018} caused by the short-range interaction. As a result, the rotation frequency associated with the $n$-th angular harmonic undergoes additional renormalization~\cite{Silin1959,Platzman1973} and is given by $\bar{\omega}_n=n(1+F_n)\omega_c$. In particular, current density rotates with the bare frequency $\bar{\omega}_1=eH/m^{(0)}c$ (main cyclotron resonance harmonic) in consistence with Kohn's theorem~\cite{Kohn1961} and the shear stress rotation is governed by $\bar{\omega}_2=2(1+F_2)\omega_c$, see Fig.~\ref{Fig:Scheme}a.

To describe the collective modes of 2DEL in magnetic field within the viscoelastic model we combine the linearized Navier-Stokes and continuity equations for the non-equilibrium quasiparticle density $\delta N ({\bf r},t)$ and current density ${\bf j}({\bf r},t)$. For the waves propagating along $Ox$ with the wave vector $q$ at frequency $\omega$, the set of equations for the Fourier transformed quantities $\delta N_{\omega q}$, ${\bf j}_{\omega q}$ reads
\begin{gather}
    -{\rm i}\omega \delta N + {\rm i} q j_x=0,
    \label{Eq:Cont}\\
    \begin{split}
    [-{\rm i}\omega+\bar{\nu}_1+\eta_s(\omega)q^2] j_x &=\\
    [\bar{\omega}_1 +& \eta_H(\omega) q^2] j_y - {\rm i} q s_p^2(q)\delta N,
    \end{split}
    \label{Eq:NS_x}\\
    [-{\rm i}\omega+\bar{\nu}_1+\eta_s(\omega)q^2] j_y=-[\bar{\omega}_1 + \eta_H(\omega) q^2] j_x.
    \label{Eq:NS_y}
\end{gather}
Here $\bar{\nu}_n=(1+F_n)\tau_n^{-1}$ denote renormalized effective scattering rates~\cite{Pines_Nozieres_2018}, and $\bar{\nu}_1\ll\bar{\nu_2}$ describes momentum relaxation due to scattering by disorder. The Eqs.~(\ref{Eq:Cont})-(\ref{Eq:NS_y}) valid at strong dipole Landau interaction $F_1\gg 1$~\cite{Alekseev2019_SC} and $qR_c\ll 1$ determine the long wavelength dispersion of the two types of waves in viscoelastic 2DEL. The last term in~(\ref{Eq:NS_x}) describes restoring force for the charge density oscillations due to quasiparticle compressibility and long-range self-consistent electrostatic field $U({\bf r},t)={\int V(|{\bf r}-{\bf r}'|)\delta N({\bf r}',t)d{\bf r}'}$. Here $V(r)$ is the 2D Coulomb interaction screened by metallic gate introduced to a 2D sample to control the carrier concentration, its Fourier transform is $V(q)=\frac{2\pi e^2}{\varkappa q}(1-e^{-2qd})$, where $\varkappa$ is the static background dielectric constant and $d$ is the distance between the gate and 2DEL. At zero magnetic field, dispersion of {\it longitudinal} (see Fig.~\ref{Fig:Scheme}b) plasmon wave $\omega_p(q)=s_p(q)q$ in 2D and its velocity $s_p^2(q)=(1+F_1)(1+\tilde{F}_0)v_F^2/2$ are determined by the renormalized~\cite{Silin1958,Silin1959,Pines_Nozieres_2018} Landau parameters $\tilde{F}_n(q)=F_n+F_c(q)\delta_{n0}$ which describe total quasiparticle interaction in the $n$-channel consisting of short-range $F_n$ and Coulomb $F_c(q)=D V(q)$ parts (here $D$ stands for the 2D density of states). In magnetic field, plasmons are hybridized with current cyclotron rotation leading to magnetoplasmon with dispersion
\begin{equation}
    \omega_{\rm mp}(q)=\sqrt{\bar{\omega}_1^2+\omega_{p}^2(q)}.
\end{equation}

\begin{figure}[t!]
    \includegraphics[width=0.49\textwidth]{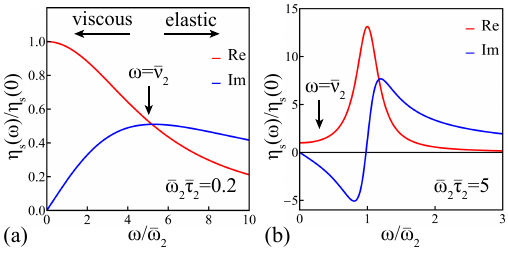}
    \caption{Viscoelasticity of 2DEL in a) weak and b) strong magnetic fields; $\bar{\omega}_2=2(1+F_2)\omega_c$ and $\bar{\nu}_2=\bar{\tau}_2^{-1}=(1+F_2)\nu_2$ denote renormalized frequencies of the shear stress cyclotron rotation and relaxation, respectively.}
    \label{Fig:VE_2DEL}
\end{figure}

Frequency-dependent kinematic shear viscosity coefficients given by
\begin{gather}
    \label{Eq:Shear_visc}
    \eta_s(\omega)={\rm i} (1+F_1)(1+F_2)\frac{v_F^2}{4}\frac{\omega+{\rm i}\bar{\nu}_2}{(\omega+{\rm i}\bar{\nu}_2)^2-\bar{\omega}_2^2},\\
    \eta_H(\omega)=(1+F_1)(1+F_2)\frac{v_F^2}{4}\frac{\bar{\omega}_2}{(\omega+{\rm i}\bar{\nu}_2)^2-\bar{\omega}_2^2},
\end{gather}
describe viscoelasticity of magnetized 2DEL. In classically weak magnetic fields $\bar{\omega}_2\bar{\tau}_2\ll 1$, viscoelastic properties of 2DEL are similar to the zero field case: Hall (odd) viscosity $\eta_H(\omega)$ is negligible and the ordinary (even) shear viscosity $\eta_s(\omega)$ becomes purely imaginary in collisionless domain $\omega\bar{\tau}_2\gg 1$ (see Fig.~\ref{Fig:VE_2DEL}a), enabling propagation of the shear sound with linear dispersion $\omega_{\rm sh}(q)=s_{\rm sh}q$. Its velocity $s_{\rm sh}^2=(1+F_1)(1+F_2)v_F^2/4$ coincides with the velocity of transverse zero sound~\cite{Landau_FL:ZS} in 2D~\cite{Khoo2019} within the applicability domain of the viscoelastic model ($F_1\gg 1$). In contrast to compression (plasmon) wave, quasiparticle density in {\it transverse} shear sound remains constant, hence its dispersion is unaffected by Coulomb interaction. Due to oscillations of non-diagonal components of the shear stress tensor $\Pi_{xy}(x,t)$ in the wave, each element of the electron liquid is deformed in direction perpendicular to the wave vector without volume change (see Fig.~\ref{Fig:Scheme}c). At classically strong magnetic fields $\bar{\omega}_2\bar{\tau}_2\ll 1$, viscous behavior is reestablished (see Fig.~\ref{Fig:VE_2DEL}b) in the part of the collisionless domain near the shear stress cyclotron rotation frequency $\bar{\omega}_2$ (see Fig.~\ref{Fig:Scheme}a). This viscoelastic resonance appearing in both $\eta_s(\omega)$ and $\eta_H(\omega)$ leads to shear magnetosound mode, representing the second solution of the system~(\ref{Eq:Cont})-(\ref{Eq:NS_y}). Its dispersion starts at the second cyclotron resonance harmonic $\omega_{\rm ms}(0)=\bar{\omega}_2$ and has two characteristic parts depending on the ratio $\bar{\omega}_1/\omega_p$,
\begin{equation}
	\label{Eq:MS_disp}
	\omega_{\rm ms}(q)=
		\begin{cases}
			\sqrt{\bar{\omega}_2^2 -2 \bar{\omega}_2 \omega_{\rm sh}^2/\bar{\omega}_1+\omega_p^2\omega_{\rm sh}^2/\bar{\omega}_1^2}, &  \bar{\omega}_1\gg \omega_p\\
			\sqrt{\bar{\omega}_2^2+\omega_{\rm sh}^2}, &  \bar{\omega}_1\ll \omega_p.
		\end{cases}
\end{equation}
Eq.~(\ref{Eq:MS_disp}) is obtained using the relations $\bar{\omega}_{2}\ll \bar{\omega}_1$ and $\omega_{\rm sh}\ll \omega_p$ between  characteristic frequencies of the viscoelastic model (provided $\tilde{F}_0,F_1\gg 1$). At $\bar{\omega}_1/\omega_p\gg 1$, longitudinal (in current polarization) plasmon and transverse shear oscillations are mixed by the circularly polarized cyclotron motion. The corresponding dispersion is characterized by negative wave velocity at small $q$ due to Hall viscosity. This effect is similar for hydrodynamic waves in neutral fluids~\cite{Alekseev2016}. At $\bar{\omega}_1/\omega_p\ll 1$, plasmon and shear waves are separated and dispersion equation takes the form $\omega=-{\rm i}\eta_s q^2$ conventional for the shear sound, but with modified viscosity~(\ref{Eq:Shear_visc}). The dispersion $\omega_{\rm ms}(q)$ in this limit was obtained in~\cite{Alekseev2019_PRL,Alekseev2019_SC}.

Both magnetoplasmon and shear magnetosound are {\it longitudinal} (with respect to self-consistent field) electrostatic modes in magnetized 2DEL. Their dispersion laws can be found from zeros of the longitudinal dielectric function $\varepsilon_{xx}(\omega{\bf ,q})=0$ calculated using Eqs.~(\ref{Eq:Cont})-({\ref{Eq:NS_y}}), while the poles of $\varepsilon_{yy}(\omega,{\bf q})$ responsible for transverse modes are absent. At short wavelengths $qR_c\sim1$, magnetoplasmon is coupled to quasiparticle cyclotron harmonics near their crossing points (see Fig.~\ref{Fig:Scheme}d), leading to formation of the conventional BMs. Longitudinal shear magnetosound generates a different type of BMs via the same mechanism. Since the threshold of the shear sound emergence in 2D $F_1\geq 1$ coincides (at $F_n=0$ for $|n|\geq 2$) with {\it inversion} of the main and the second cyclotron resonance harmonics $\bar{\omega}_2\leq \bar{\omega}_1$ (see Fig.~\ref{Fig:Scheme}d), magnetoplasmon and shear magnetosound dispersions do not cross and the latter forms its own series of BMs independently.


\section{Bernstein modes in 2DEL}
To obtain dispersion of BMs in 2DEL we go beyond the viscoelastic model and use Landau-Silin kinetic equation~\cite{Silin1958,Platzman1973,Pines_Nozieres_2018} valid in the quasiclassical domain at arbitrary strength of the short-range interaction. Its linearized form is
\begin{equation}
	\partial_t \delta n + [{\bf v}_F {\bm \nabla}+\omega_c \partial_{\theta}]\delta \bar{n}-{\bf v}_F[\bm{\nabla}U+e{\bf E}_0]n_0'={\rm St}[\delta\bar{n}],
	\label{Eq:LS_KinEq}
\end{equation}
where ${\bf E}_0$ is the external electric field, $\theta$ is the angle between quasiparticle velocity at the Fermi surface and $Ox$ (the geometry is shown in Fig.~\ref{Fig:Scheme}). Here $\delta n_{\bf p}({\bf r},t)$ stands for the global non-equilibrium part of distribution function, $\delta\bar{n}_{\bf p}({\bf r},t)=n_{\bf p}({\bf r},t)-n_0(\bar{\mathcal{E}}_{\bf p}({\bf r},t))$ describes deviation from the local equilibrium~\cite{Pines_Nozieres_2018} characterized by the local energy of quasiparticle $\bar{\mathcal{E}}_{\bf p}({\bf r},t)=\epsilon({\bf p})+\delta \mathcal{E}_{\bf p}({\bf r},t)$ consisting of the non-interacting part $\epsilon({\bf p})=p^2/2m$ and the contribution from Landau short-range interaction $\delta\mathcal{E}_{\bf p}({\bf r},t)=\sum_{{\bf p}'}f_s(\widehat{{\bf p}{\bf p}'})\delta n_{{\bf p}'}({\bf r},t)$. We describe weakly excited state of 2DEL in terms of the small Fermi surface deformation $\Phi(\theta,{\bf r},t)\ll \mu$ related to the distribution function via $\delta n_{\bf p}({\bf r},t)=-n_0'(\epsilon)\Phi(\theta,{\bf r},t)$, where $n_0(\epsilon)=\Theta(\mu-\epsilon)$ is the equilibrium distribution and $\mu$ is the Fermi energy.

We consider plane-wave like (${\bf q}||Ox$) solutions $\Phi(\theta,{\bf r},t)=\Phi(\theta)e^{{\rm i}({\bf qr}-\omega t)}$ of the homogeneous form of kinetic equation~(\ref{Eq:LS_KinEq}) in collisionless regime, thus we omit the collision integral ${\rm St}[\delta\bar{n}]$ everywhere, except for regularization of singularities. Expanding Fermi surface deformation $\Phi(\theta)=\sum_n \Phi_n e^{{\rm i}n\theta}$ and spin-symmetric part of the Landau interaction function $f_s(\widehat{{\bf p}_F{\bf p}_F'})=D^{-1}\sum_n F_n e^{{\rm i} n (\theta-\theta')}$ in Fourier series (note that $F_n=F_{-n}$) and combining them with Eq.~(\ref{Eq:LS_KinEq}), we obtain closed system for the angular harmonics of $\Phi(\theta)$
\begin{gather}
    \label{Eq:Phi_syst}
    (1+\tilde{F}_n)\Phi_n-\sum\limits_s A_{ns}\tilde{F}_s\Phi_s=0,\\
    A_{ns}(\Omega,k)=\Omega\sum\limits_{l}\frac{J_{l-n}(k)J_{l-s}(k)}{\Omega-l+{\rm i}\gamma},
 \end{gather}
where $\Omega=\omega/\omega_c$ and $k=qR_c$ are the dimensionless frequency and wave vector, $J_n(k)$ is the Bessel function of the first kind and $\gamma=\nu_{ee}/\omega_c$ stands for the dimensionless quasiparticle scattering rate appearing in the model collision integral ${\rm St}[\delta\bar{n}]=n_0'\nu_{ee}\sum_{|n|\geq 2} \Phi_n e^{{\rm i} n \theta}$, which we use in this work. At zero wave vector, the solutions of Eq.~(\ref{Eq:Phi_syst}) are ${\rm Re}\,\omega(0)=\bar{\omega}_n$. 

\begin{figure}[t!]
    \includegraphics[width=0.49\textwidth]{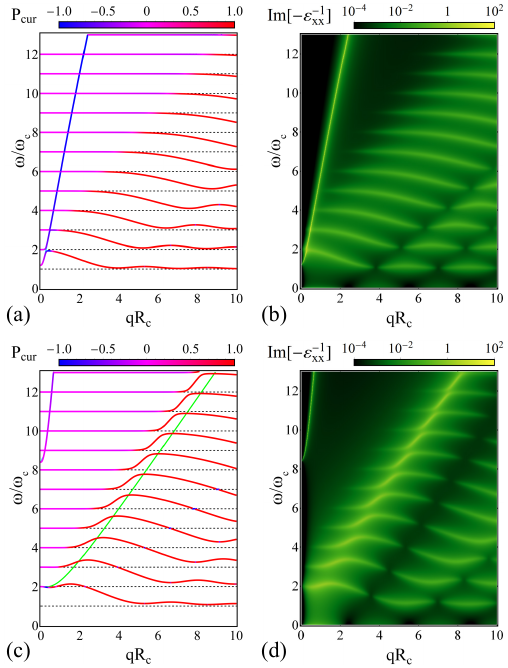}
    \caption{Bernstein modes spectrum (a,c) and its visualization through the energy loss function (b,d) in gated 2DEL with $R_c/a_B=300$, $d/R_c=0.1$ for weak $F_0=-0.4$, $F_1=0.2$ (a,b) and strong $F_0=-2$, $F_1=7.4$ (c,d) dipole Landau interaction at $\gamma=0.1$. Color of the dispersion curves in a) and c) show linear current polarization degree. Cyan and light green curves denote long wavelength dispersions of magnetoplasmon and shear magnetosound, respectively.}
    \label{Fig:BM_scrn}
\end{figure}


In this work, we consider the effect of short-range interaction in isotropic and dipole channels only~\cite{Landau_FL:ZS,Silin1959,Khoo2019,Klein2019,Khoo2020,Valentinis2021,Valentinis2021_SciRep}, thus the system~(\ref{Eq:Phi_syst}) reduces to
\begin{equation}
    \label{Eq:BM_syst}
    \begin{bmatrix}
        1- \beta A_{11}^{+} & -\tilde{F}_0 A_{10} & 1- \beta A_{11}^{-}\\
        - \beta A_{01}^{+} & 1+\tilde{F}_0(1- A_{00}) & -\beta A_{01}^{-}\\
        1-\beta A_{\bar{1}\bar{1}}^{+} & -\tilde{F}_0 A_{\bar{1}0} & -1+\beta A^{-}_{\bar{1}\bar{1}}
    \end{bmatrix}
    \begin{bmatrix}
        j_{x}\\
        v_F\delta N\\
        {\rm i} j_{y}
    \end{bmatrix}
    =0,
\end{equation}
where  $A_{ns}^{\pm}=A_{ns}\pm A_{n\bar{s}}$, $\bar{s}=-s$, $\beta=F_1/(1+F_1)$ and non-equilibrium quasiparticle and current densities are given by $\delta N({\bf r},t)=2\sum_{\bf p}\delta n_{\bf p}({\bf r},t)$ and ${\bf j}({\bf r},t)=2\sum_{\bf p} {\bf v}_{\bf p}\delta\bar{n}_{\bf p}({\bf r},t)$. Eqs.~(\ref{Eq:BM_syst}) are analogous to macroscopic equations of the viscoelastic model~(\ref{Eq:Cont})-({\ref{Eq:NS_y}}) valid at short wavelengths and arbitrary short-range interaction strength. In 2DEL, isotropic Landau parameter $F_0$ is typically negative and decreases with $r_s$~\cite{Eisenstein1992,Eisenstein1994,Giuliani_Vignale_2005}, while the dipole one $F_1$ is positive and increases with $r_s$~\cite{Shashkin2002,Pudalov2002,Anissimova2006,Gokmen2010,Kuntsevich2015,Falson2015,Melnikov2023}. 


\begin{figure}[t!]
    \includegraphics[width=0.49\textwidth]{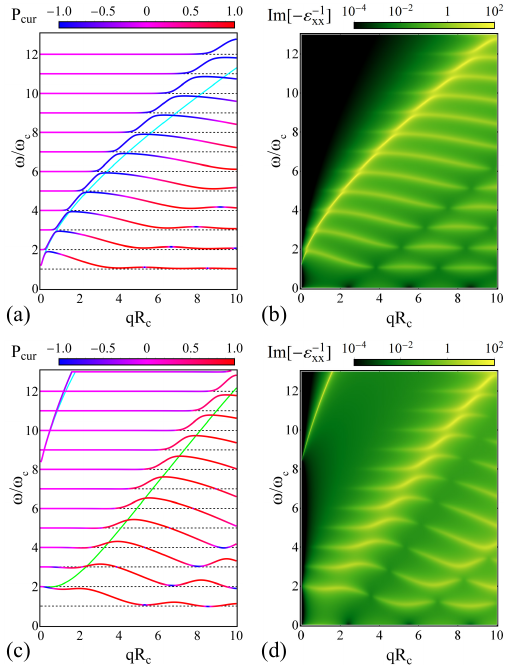}
    \caption{Bernstein modes spectrum (a,c) and its visualization through the energy loss function (b,d) in ungated 2DEL with $R_c/a_B=15$, $d/R_c=10$, for weak $F_0=-0.4$, $F_1=0.2$ (a,b) and strong $F_0=-2$, $F_1=7.4$ (c,d) dipole Landau interaction at $\gamma=0.1$. Other notations are the same as in Fig.~\ref{Fig:BM_scrn}.}
    \label{Fig:BM_bare}
\end{figure}


The BM spectra of the gated and ungated 2DELs with either weak and strong dipole Landau interaction calculated after Eqs.~(\ref{Eq:BM_syst}) are shown in Fig.~\ref{Fig:BM_scrn}a,c and Fig.~\ref{Fig:BM_bare}a,c, respectively. The color of the curves corresponding to the linear current polarization degree $P_{\rm cur}=(|j_y|^2-|j_x|^2)/(|j_y|^2+|j_x|^2)$ visualizes nature of the modes (at particular $q$): 1) plasmon-like ($P_{\rm cur}=-1$) with longitudinal current polarization, 2) shear wave like ($P_{\rm cur}=1$) carrying the transverse current or 3) cyclotron motion like ($P_{\rm cur}=0$) with circular current polarization. In the weakly interacting case (Figs.~\ref{Fig:BM_scrn}a and~\ref{Fig:BM_bare}a) the conventional BM pattern is reproduced and the magnetoplasmon-like segments of dispersion are characterized by longitudinal current polarization. With the increase of $F_1$, magnetoplasmon dispersion and the associated BMs are shifted upwards relative to quasiparticle harmonics. At strong dipole interaction, the calculated spectra (Figs.~\ref{Fig:BM_scrn}c and~\ref{Fig:BM_bare}c) demonstrate inversion of the main and the second cyclotron resonance harmonics and reconstruction of BM pattern regardless of the screening strength (induced by the gate). Shear BMs with transverse current polarization arise due to anti-crossings of shear magnetosound dispersion and quasiparticle cyclotron harmonics. In contrast to the conventional BM case, each branch of the new type spreads to a wider frequency range greater than $\omega_c$.

In Fig.~\ref{Fig:BM_scrn}b,d and Fig.~\ref{Fig:BM_bare}b,d we plot maps of the energy loss function ${\rm Im}\,[-\varepsilon_{xx}^{-1}(\omega,{\bf q})]$~\cite{Pines_Nozieres_2018,Platzman1973} (determined by the longitudinal nonlocal proper conductivity calculated using the full kinetic equation~(\ref{Eq:LS_KinEq})), which describes the inelastic scattering of high energy electrons passing through the system~\cite{Landau_V8}. Positions of the peaks of ${\rm Im}\,[-\varepsilon_{xx}^{-1}(\omega,{\bf q})]$ on the $(\omega,q)$ plane and their widths correspond to dispersion and damping of the longitudinal collective modes, respectively. It is well seen that the shear BMs are well-defined and directly accessible through the energy-loss techniques.


\section{Conclusions}
In this work we have considered the emergence of shear BMs and inversion of the cyclotron resonance harmonics due to strong dipole Landau interaction. Sensitivity of BM spectrum to the short-range quasiparticle interaction can be used for spectroscopy of 2DEL. Inversion of the cyclotron resonance frequencies and emergence of shear BMs can be treated as a hallmark of strong interaction in 2DEL. Similarly to the shear magnetosound~\cite{Alekseev2019_PRL, Du2022,Alekseev_Alekseeva_2022}, shear BMs can be excited in the ac magnetotransport experiments and may lead to the giant peak in photoresistance at second and higher cyclotron harmonics. The effect may be used for estimation of the anisotropic part of Landau interaction function at various $r_s$ and magnetic fields, by analogy with the 3D metal case~\cite{Platzman1973} where excitation of the cyclotron waves is involved. Interestingly, reconstruction of the BM spectrum can happen in higher angular channels at $F_n\gg 1$, and emergence of the modes associated with the corresponding zero sound branch (e.g. for quadrupole interaction see~\cite{Silin1959,Klein2019,Aquino2019,Aquino2020}) is expected, as well as cyclotron resonance frequencies inversion with respect to the $n$-th harmonic $\bar{\omega}_n$.

\begin{acknowledgments}
A.N.A. and P.S.A. are grateful to the Foundation for the Advancement of Theoretical Physics and Mathematics ``BASIS'' (Grant No. 23-1-2-25-2) for financial support of analytical calculations based on kinetic equation in Section 3. The work of A.N.A. (numerical calculations in Section 3) was supported by the Ministry of Science and Higher Education of the Russian Federation [project no. 075-15-2020-790].
\end{acknowledgments}

\bibliography{CWEL_bib}

\end{document}